\documentclass[conference]{IEEEtran}

\usepackage{amsmath,amssymb,amsfonts}
\usepackage{algorithmic}
\usepackage{graphicx}
\usepackage{textcomp}
\usepackage{xcolor}
\usepackage{subcaption}
\usepackage{multirow}
\usepackage{array}
\usepackage{tabu}
\usepackage{siunitx}
\usepackage{adjustbox}
\usepackage{cleveref}
\usepackage{cite}
\usepackage{setspace}
\usepackage{adjustbox}
\usepackage{units}
\def\BibTeX{{\rm B\kern-.05em{\sc i\kern-.025em b}\kern-.08em
    T\kern-.1667em\lower.7ex\hbox{E}\kern-.125emX}}

\begin{document}

\title{SELD-TCN: Sound Event Localization \& Detection via Temporal Convolutional  Networks}

\author{\IEEEauthorblockN{Karim Guirguis$^{1,2}$, Christoph Schorn$^{1}$, Andre Guntoro$^{1}$, Sherif Abdulatif$^{2}$, Bin Yang$^{2}$}
\IEEEauthorblockA{\textit{$^{1}$Robert Bosch GmbH, Renningen, Germany}\\ \textit{$^{2}$University of Stuttgart, Institute of Signal Processing and System Theory, Stuttgart, Germany}\\
fixed-term.karim.guirguis, christoph.schorn,  andre.guntoro@de.bosch.com}
}

\maketitle

\begin{abstract}
The understanding of the surrounding environment plays a critical role in autonomous robotic systems, such as self-driving cars. Extensive research has been carried out concerning visual perception. Yet, to obtain a more complete perception of the environment, autonomous systems of the future should also take acoustic information into account. Recent sound event localization and detection (SELD) frameworks utilize convolutional recurrent neural networks (CRNNs). However, considering the recurrent nature of CRNNs, it becomes challenging to implement them efficiently on embedded hardware. Not only are their computations strenuous to parallelize, but they also require high memory bandwidth and large memory buffers. In this work, we develop a more robust and hardware-friendly novel architecture based on a temporal convolutional network (TCN). The proposed framework (SELD-TCN) outperforms the state-of-the-art SELDnet performance on four different datasets. Moreover, SELD-TCN achieves 4x faster training time per epoch and 40x faster inference time on an ordinary graphics processing unit (GPU).
     
\end{abstract}


\begin{figure*}[t!]
	\def\svgwidth{0.89\textwidth}
	\centering
	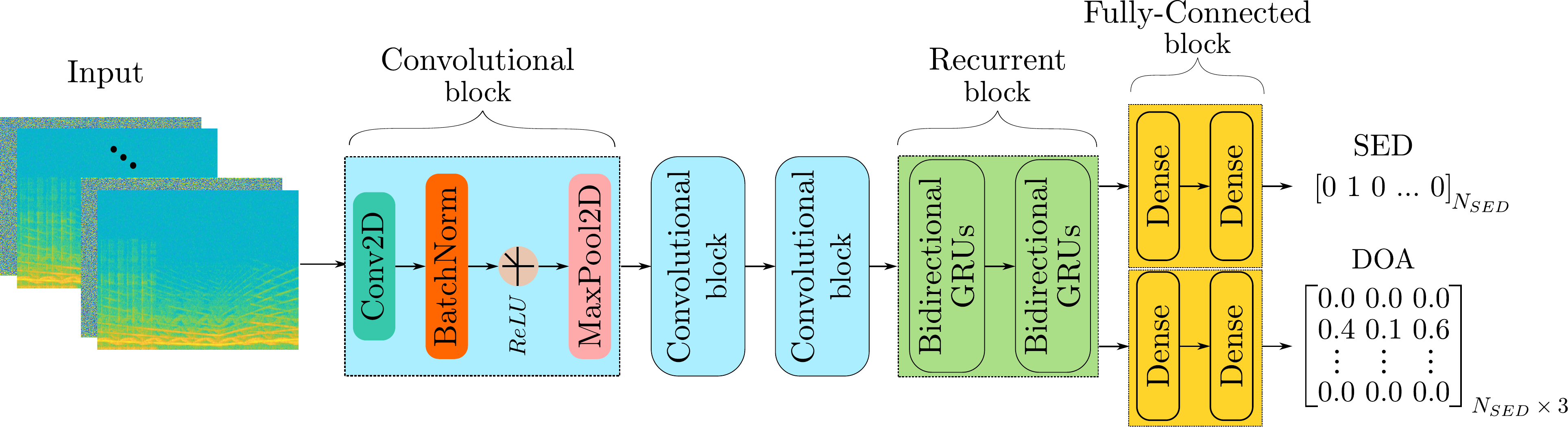
	\caption{An overview of the SELDnet architecture}
	\label{fig:seldnet}
\end{figure*}

\section{Introduction}
Sound event localization and detection (SELD) is the task of jointly identifying the temporal and spatial location of each sound event in a given audio recording. Furthermore, SELD can be broken down into two separate tasks: sound event detection (SED) and the direction of arrival (DOA) estimation. SED is a multi-label classification problem of recognizing both the onset and offset time of a sound event in a given audio recording. It is widely used in wildlife monitoring \cite{wildlife}, music \cite{musictrans,musicinst}, and automatic audio indexing \cite{autoaudioidx}. DOA estimation \cite{doa1, doa2, doa3, doa4, doa5, doa6, doanet} depicts a regression task of identifying the spatial location of the active sound events. In the automotive context, understanding the vehicle’s surrounding acoustic environment (e.g. emergency vehicles, honking) can be of high significance. Not only can SELD complement the other sensor modalities and improve safety, but it also can be realized in a more power-efficient way than visual perception. Furthermore, SELD can be categorized into \textit{monophonic} and \textit{polyphonic}. Monophonic systems recognize at most one sound event for any instance of time. However, polyphonic systems can detect multiple overlapping sound events at any instance of time.

Early SED approaches \cite{classic1,classic2} were mostly based on Gaussian mixture models (GMM) and hidden Markov models (HMM), or combination of both. The aforementioned models act as classifiers with Mel frequency cepstral coefficients (MFCCs) as the input features. However, this approach typically delivers unsatisfactory performance and requires manual parametrization, e.g. the number of simultaneously active events, to perform polyphonic scenario. Hence, non-negative matrix factorization (NMF) \cite{nmf1, nmf2, nmf3, nmf4} has been used to learn a dictionary of basis vectors from the given time-frequency (TF) spectrogram. Nevertheless, the NMF-based methods process the audio tracks in a frame-by-frame manner neglecting the correlation in time. Thus, extensive post-processing like temporal smoothing is required to achieve better temporal stability. 

Recently, deep neural networks (DNN) have been considered for SED and DOA estimation tasks using convolutional neural networks (CNNs) \cite{sedcnn1, sedcnn2, sedcnn3},  recurrent neural networks (RNNs) \cite{sedrnn1, sedrnn2, sedrnn3}, or convolutional recurrent neural networks (CRNNs) \cite{sedcrnn1, sedcrnn2, sedcrnn3, sedcrnn4, sedcrnn5, sedcrnn6}. RNNs are an essential component of various sequence modeling applications (e.g., language modeling and machine translation). They propagate a vector of hidden activations, which encapsulates the learned information through time. However, the realization of RNNs on embedded hardware imposes some challenges, including high memory requirements and power consumption \cite{rnnhw1,rnnhw2}. 

CNNs have been enabling vast advancements in various computer vision tasks, e.g., image classification, semantic segmentation, and object detection \cite{cnns}. CNNs act as an automatic feature extractor by learning a set of kernels. A fixed-size kernel slides over the input image computing the dot product on each sub-block resulting in a feature map. The deeper the CNNs, the more complex features it learns through a non-linear combination of the previous simpler ones. Unlike RNNs, the nature of the convolution operation makes CNN highly parallelizable and easier to train. 


To capture long-term dependencies, a new family of CNNs have been introduced, namely temporal convolutional networks (TCNs) \cite{wavetcn, tcn}. Rather than normal convolutions, TCNs utilize dilated convolutions to enlarge their receptive field. This allows a wider part of the input data to contribute to the output. TCNs made their first appearance in the WaveNet \cite{wavetcn}. Compared to RNNs, TCNs can be of great advantage for numerous reasons. Firstly, rather than sequentially processing an input sequence having to wait for the output from previous time steps, TCNs process the whole sequence in parallel via convolutions. Furthermore, receptive field size can be flexibly changed through increasing their dilation rate or kernel sizes. Moreover, the utilization of residual blocks along with skip connections provides a path through which gradients can easily flow, resulting in stable gradients during training.

 
In this paper, we investigate the SELD performance when replacing the bidirectional RNNs with TCNs. Additionally, we study the robustness of both models on different types of noise and reverberance . To our knowledge, this is the first attempt to perform a SELD task via TCNs. The proposed framework reformulates the state-of-the-art SELDnet \cite{seldt} by utilizing a TCN-based architecture instead of conventional bidirectional RNNs. Employing a convolutional-based architecture would be a significant step towards a hardware-friendly implementation. This, in turn, allows the use of a single hardware engine for the whole network. In section \ref{methods}, the current state-of-the-art SELDnet model will be introduced in detail, followed by our proposed architecture. Comparisons were conducted to investigate the performance of the proposed SELD-TCN against the SELDnet.

\section{Methodologies}\label{methods}
\subsection{SELDnet:}
An overview of the baseline architecture, SELDnet, is shown in Fig. \ref{fig:seldnet}. 
\paragraph{Input representation}
Given a multichannel audio recording, sampled at \unit[44.1]{kHz}, the spectrogram is extracted for each channel via a short-time Fourier transform (STFT). Both the phase and magnitude of the spectrogram are stacked channel-wise as separate input features. In our implementation, we use a hamming window of length $512$ samples with a $50\%$ overlap.   
\paragraph{Convolutional block}
In order to learn both inter-/ and intra-channel features, the input is fed through three consecutive convolutional blocks. All convolutions are parameterized with 64 filters of size $3 \times 3$ and a $1 \times 8$ max pooling with an exception of $1 \times 2$ for the final layer. Dimensionality reduction is performed along the frequency axis to maintain the temporal dimension unchanged. Moreover, the output activations are normalized after each CNN via batch normalization \cite{batchnorm} followed by a rectified linear unit (ReLU) activation.
\paragraph{Recurrent block} 
To learn the temporal context information, output activations from the convolutional block are reshaped into 1D then fed to two consecutive bidirectional RNN layers. In contrast to unidirectional RNNs, the bidirectional structure makes it possible to incorporate future knowledge. As a result, bidirectional networks can exploit the full context of an input sequence. Gated recurrent units (GRUs) are modified architectures of RNNs featuring an update gate and reset gate. Intuitively, GRUs try to learn which information worth saving and which is unnecessary that can be removed. This, in turn, helps to eliminate the vanishing gradient problem which comes with the normal RNNs. Each RNN layer contains a sequence of $128$ GRU cells with tanh activations. 
\paragraph{Fully-connected block}
The output of the recurrent layers is fed to two parallel branches of fully-connected blocks. Each consists of two full-connected (FC) layers, where the first layer in both branches has $N_{FC} = 128$ neurons. The SED branch has a second layer with number of neurons equal to that of the classes $N_{SED}$, representing the number of known events in the corresponding dataset. For the  DOA estimation, the second layer has $3 \times N_{SED}$ neurons, each denoting the 3D Cartesian coordinates at the origin of a unit sphere around the microphone array. The final layer in SED branch utilizes a sigmoid activation, to ensure the output range $[0, 1]$, with 1 being for active sound event. On the other hand, DOA estimation branch apply a tanh activation for an output range $[-1, 1]$ representing the $x$, $y$, and $z$ coordinates, respectively.

\begin{figure*}[t!]
	\hspace{-3cm}
	\centering
	\def\svgwidth{\textwidth}
	\begin{subfigure}[b]{0.3\textwidth}
		\centering
		\def\svgwidth{0.82\textwidth}
		\subcaptionbox{\label{fig:seldtcn}SELD-TCN architecture}{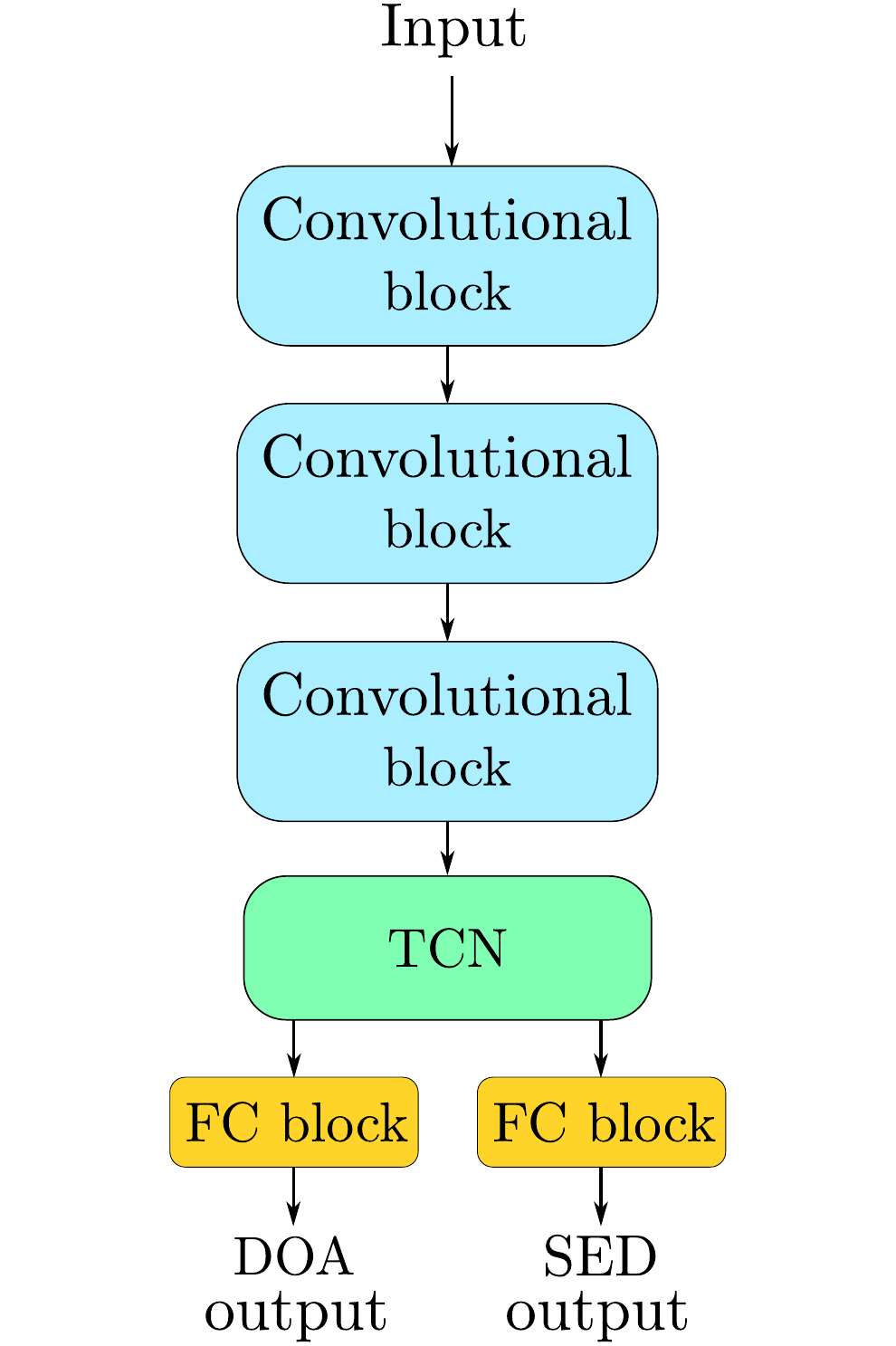}
	\end{subfigure}
	\hspace{0.5cm}
	\begin{subfigure}[b]{0.3\textwidth}
		\centering
		\def\svgwidth{\textwidth}
		\subcaptionbox{\label{fig:tcn}TCN block architecture}{\hspace{-0.8cm}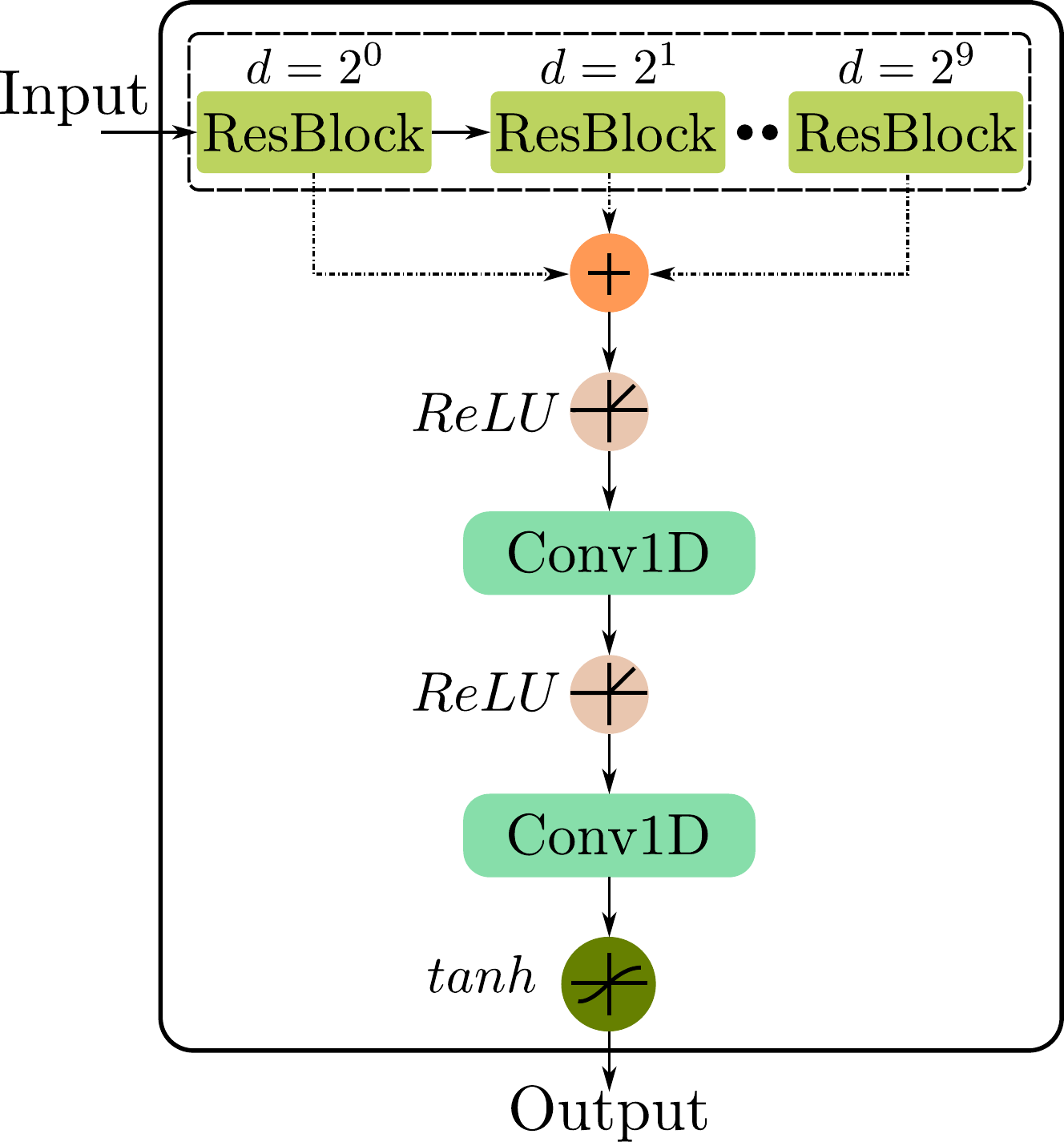}
	\end{subfigure}
	\hspace{-0.5cm}
	\begin{subfigure}[b]{0.3\textwidth}
		\centering
		\def\svgwidth{1.15\textwidth}
		\subcaptionbox{\label{fig:resblock}ResBlock architecture}{\hspace{1.2cm}\vspace{-0.1cm}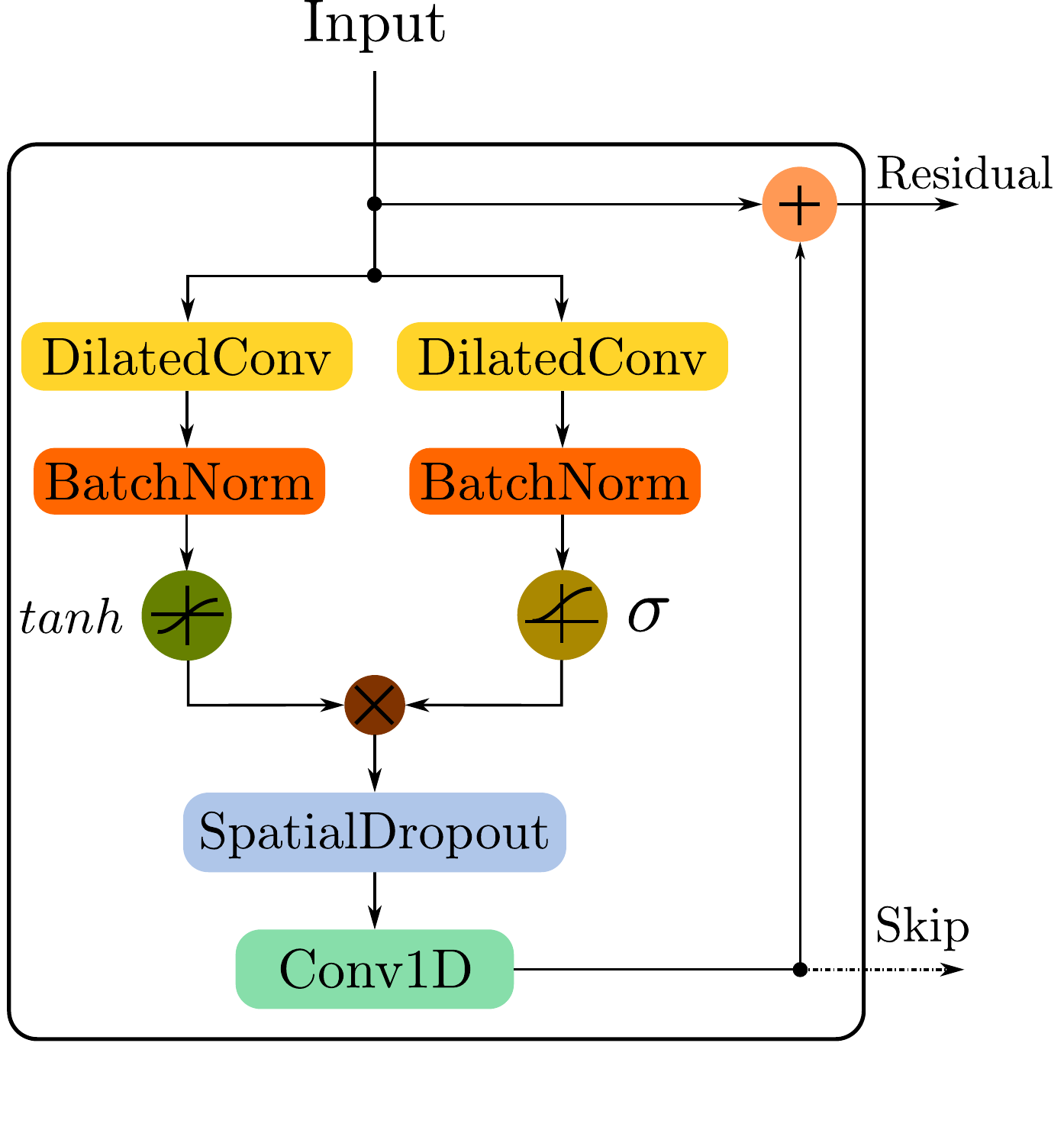}
	\end{subfigure}
	
	\captionsetup{justification=centering}
	\caption{\label{fig:seldtcnarchi}An overview of the proposed SELD-TCN architecture}
	\vspace{-3mm}
\end{figure*}


\subsection{SELD-TCN:}
%
An overview of our proposed SELD-TCN architecture is shown in Fig. \ref{fig:seldtcnarchi}. Unlike RNNs, TCNs are trying to learn the underlying structure in a sequence by employing dilated convolutions that enable TCNs to process a wider range of the input. When $d = 1$ a normal convolution is performed. On the other hand, the higher the dilation factor the larger receptive fields are exploited, i.e. expanding the kernel over more comprehensive input regions.  For TCNs to achieve such exponentially large receptive fields, it requires a deeper network. Consequently, they can suffer from unstable gradients during backpropagation. In an attempt to overcome this limitation, residual connections are utilized to learn modifications to the identity mapping rather than the transformations itself.  
The aforementioned SELDnet architecture is used while replacing the recurrent block by a TCN, where we adopt the WaveNet \cite{wavenetdenoise} architecture. Fig. \ref{fig:tcn} shows an overview of the proposed TCN block architecture. Normally, TCNs utilize causal convolutions, where the output at time $t$ depends solely on the current and past elements. This can be achieved by zero paddings. However, to mimic the bidirectional RNNs' use of future knowledge, we modify all convolutions within the TCN block to be non-causal. In this implementation, we use $10$ residual blocks (ResBlocks) with dilation rates $d=\{ 2^0, 2^1, .., 2^9 \}$. As shown in Fig. \ref{fig:resblock}, the ResBlock consists of 1D dilated convolutions with $256$ filters of size $3$ and a dilation rate $d$. Unlike the original proposed ResBlock in \cite{wavenetdenoise}, we employ batch normalization right after the dilated convolutions. Next, the normalized activations are passed to tanh and sigmoid activation, respectively. Prior to the final convolution layer, featuring $256$ filters of size $1$, we utilize a spatial dropout with a dropout rate of $0.5$. The output activations will be output as a skip connection and also summed up with the original input forming the residual connection. Next, a ReLU activation is applied to the summation of all skip connections. Finally, the output activations is fed to two consecutive convolution layers, where each consists of $128$ filters of size $1$.  
Utilizing TCNs usually comes at the cost of a larger number of parameters and multiply-and-accumulate (MAC) operations. However, the advances in hardware accelerators enable the realization of such architectures while reducing the energy cost, silicon area, and memory size requirements \cite{hwdnnsurvey}. Given the nature of CNNs, accelerators strive to harness three various sorts of data reuse. Firstly, the convolutional reuse, where the same filter map activations and filter weights are utilized within a provided channel. Secondly, since the same input feature map is convoluted with multiple filters, the input feature map activations can be reused in addition. Finally, the identical filter weights are employed across the input feature maps when performing batch processing. On the other hand, RNNs acceleration faces two main challenges: Firstly, it cannot fully exploit parallelism due to its recurrent nature involving serial processing. The second challenge is the data dependency of computations. Hence, a customized accelerator has to be designed while being particularly tailored to its computation model.

\begin{table*}
	\caption{Methods evaluation on different sampling rates.\vspace{-2mm}}
	\label{tab:sampling}
	\centering
	\Large
	\bgroup
	\def\arraystretch{1.1}
	\begin{adjustbox}{width=1.0\textwidth}
		\begin{tabular}{c|c|cccc|cccc|cccc|cccc}
			\noalign{\smallskip} \hline \hline \noalign{\smallskip}
			\multirow{2}{*}{Method} & \multirow{2}{*}{Sampling Rate} & \multicolumn{4}{c}{ANSYN} & %
			\multicolumn{4}{c}{MANSYN} & \multicolumn{4}{c}{REAL} & \multicolumn{4}{c}{MREAL}\\
			& & F1 & ER & FR & DE & F1 & ER & FR & DE & F1 & ER & FR & DE & F1 & ER & FR & DE\\
			\hline
			SELDnet &  \multirow{2}{*}{44.1 [\si{kHz}]} & 93.4 & 0.11 & 81.2 & 17.5 &
			94.6 & 0.10 & 90.7 & 14.2 &
			74.1 & 0.39 & 48.2 & 38.1 &
			70.5 & 0.44 & 46.2 & \textbf{41.2}\\
			
			SELD-TCN & & \textbf{95.5} & \textbf{0.08} & \textbf{86.8} & \textbf{16.0} &
			\textbf{95.4} & \textbf{0.09} & \textbf{92.7} & \textbf{13.5} &
			\textbf{75.1} & \textbf{0.39} & \textbf{52.4} & \textbf{35.8} &
			\textbf{72.2} & \textbf{0.42} & \textbf{46.6} & 41.7 \\		
			\hline
			SELDnet &  \multirow{2}{*}{16 [\si{kHz}]} & 94.6 & 0.09 & 83.1 & 16.0 &
			95.8 & 0.07 & 90.4 & 14.0&
			76.1 & 0.36 & 51.6 & 36.5 &
			71.6 & 0.42 & 46.6 & \textbf{42.7}\\
			
			SELD-TCN & & \textbf{96.0} & \textbf{0.07} & \textbf{86.6} & \textbf{15.7} &
			\textbf{96.0} & \textbf{0.06} & \textbf{91.5} & \textbf{13.7} &
			\textbf{79.7} & \textbf{0.32} & \textbf{55.9} & \textbf{34.0} &
			\textbf{73.1} & \textbf{0.40} & \textbf{46.6} & 43.9 \\	
			\hline
			SELDnet &  \multirow{2}{*}{8 [\si{kHz}]} & 93.1 & 0.12 & 76.9 & 19.9 &
			95.4 & 0.08 & 87.8 & 16.6 &
			75.0 & 0.37 & 52.1 & 38.6 &
			72.2 & 0.40 & 45.7 & \textbf{42.8}\\
			
			SELD-TCN & & \textbf{95.2} &\textbf{ 0.08} & \textbf{82.3} & \textbf{18.1} &
			\textbf{95.5} & \textbf{0.08} & \textbf{88.6} & \textbf{16.3} &
			\textbf{77.8} &\textbf{ 0.34} & \textbf{53.5} & \textbf{37.3} &
			\textbf{72.9} & \textbf{0.39} & \textbf{47.5} & 44.5 \\		
			\noalign{\smallskip} \hline \noalign{\smallskip}
		\end{tabular}
	\end{adjustbox}
	\egroup
	\vspace{-2mm}
\end{table*}

\begin{table*}
	\caption{Quantitative comparisons on the downsampled 16kHz noisy datasets. \vspace{-2mm}}
	\label{tab:noise}
	\centering
	\Large
	\bgroup
	\def\arraystretch{1.1}
	\begin{adjustbox}{width=1.0\textwidth}
		\begin{tabular}{c|c|cccc|cccc|cccc|cccc}
			\noalign{\smallskip} \hline \hline \noalign{\smallskip}
			\multirow{2}{*}{Method} & \multirow{2}{*}{Description} & \multicolumn{4}{c}{ANSYN} & %
			\multicolumn{4}{c}{MANSYN} & \multicolumn{4}{c}{REAL} & \multicolumn{4}{c}{MREAL}\\
			& & F1 & ER & FR & DE & F1 & ER & FR & DE & F1 & ER & FR & DE & F1 & ER & FR & DE\\
			\hline
			SELDnet &  \multirow{2}{*}{AWGN} & 93.8 & 0.10 & 80.0 & \textbf{16.5} &
			93.8 & 0.10 & 84.0 & 17.2 &
			72.2 & 0.40 & 47.3 & 40.9 &
			73.9 & 0.37 & 48.2 & 40.0\\
			
			SELD-TCN & & \textbf{94.3} & \textbf{0.09} & \textbf{82.6} & 16.8 &
			\textbf{94.7} & \textbf{0.08} & \textbf{87.2} & \textbf{15.7} &
			\textbf{77.7} & \textbf{0.35} & \textbf{52.7} & \textbf{34.9} &
			\textbf{77.6} & \textbf{0.34} & \textbf{50.9} & \textbf{37.8} \\
			
			\hline
			SELDnet &  \multirow{2}{*}{STREET} & 93.4 & 0.11 & 80.0 & \textbf{16.1} &
			94.2 & 0.09 & 85.4 & 16.1 &
			64.1 & 0.48 & 39.8 & 43.8 &
			65.3 & 0.47 & 40.1 & 44.2\\
			
			SELD-TCN & & \textbf{94.5} & \textbf{0.09} & \textbf{82.2} & 16.6 &
			\textbf{95.4} & \textbf{0.07} & \textbf{88.6} & \textbf{14.5} &
			\textbf{74.7} & \textbf{0.38} & \textbf{50.0} & \textbf{36.5} &
			\textbf{69.9} & \textbf{0.44} & \textbf{ 46.7} & \textbf{42.1} \\	
			
			\hline
			SELDnet &  \multirow{2}{*}{REVERB} & 93.8 & 0.10 & 79.7 & \textbf{16.5} &
			93.9 & 0.10 & 84.8 & 17.0 &
			70.9 & 0.41 & 46.9 & 39.0 &
			66.2 & 0.47 & 43.0 & 43.7\\
			
			SELD-TCN & & \textbf{94.9} & \textbf{0.08} & \textbf{83.8} & 17.0 &
			\textbf{95.2} & \textbf{0.08} & \textbf{87.3} & \textbf{16.7} &
			\textbf{79.9} & \textbf{0.32} & \textbf{55.1} & \textbf{33.1} &
			\textbf{75.9} & \textbf{0.38} & \textbf{50.7} & \textbf{39.3} \\	
			
			\noalign{\smallskip} \hline \noalign{\smallskip}
		\end{tabular}
	\end{adjustbox}
	\egroup
	\vspace{-2mm}
\end{table*}

\section{Evaluation}\label{eval}
\subsection{Datasets}
The proposed SELD-TCN is evaluated on the Tampere University of Technology (TUT) Sound Events 2018 dataset \cite{dataset}. All experiments were performed employing four diverse sub-datasets (ANSYN, MANSYN, REAL, and MREAL) with a maximum of three temporally overlapping sound events. Each dataset contains 900, \unit[30]{s} long, first-order ambisonics (FOA), i.e. four-channel recordings with a 44.1kHz sampling rate. The datasets were subdivided into 60\%-20\%-20\% splits for training, validation, and testing, respectively. 

ANSYN and MANSYN datasets are synthesized with artificial impulse responses (IR) in an anechoic sound scene. Each contains 11 sound event classes with azimuth and elevation ranging $[-60 \si{\degree}, 60 \si{\degree})$ placed in a 10\si{\degree} resolution spatial grid. On the other hand, REAL and MREAL are synthesized with real-life IR featuring 8 sound event classes with azimuth and elevation ranging $[-40 \si{\degree}, 40 \si{\degree})$. Both MANSYN and MREAL datasets contain a moving sound source instead. 
\subsection{Metrics}
Performance of both the SELDnet and SELD-TCN is evaluated using the proposed metrics in \cite{seldnet}. Analogous to standard polyphonic SED metrics, F-score (F1) and error rate (ER) across segments of one second without overlap are utilized. F1 is defined as the harmonic mean of the precision and recall. Moreover, ER depicts the amount of errors with regard to segment-wise substitutions, insertions, and deletions. Ideally, the method will have $F1 = 1$ and $ER=0$. For localization/tracking, performance is evaluated via frame recall (FR) and DOA error (DE). FR denotes the number of predicted DOAs per frame matching the ground truth. DE is the average error between the predicted and estimated DOAs.

\subsection{Experiments} 
To evaluate the performance of the proposed SELD-TCN, we have conducted two sets of experiments. The objective of the first experiment was twofold: firstly, to investigate the performance of the SELD-TCN on the four aforementioned datasets, thus establishing a fair comparison against the SELDnet. Secondly, to examine the robustness of both models against reduced sampling frequencies (44.1, 16, and \unit[8]{kHz}).

The goal of the second experiment was to evaluate the robustness of both models against noise and reverberance. For simplicity, we chose to perform such an investigation on the \unit[16]{kHz} downsampled version of the datasets. Two different types of noise were applied: additive white Gaussian noise (AWGN) and multi-channel street noise from the DEMAND \cite{demand} corpus (STREET). Different SNR levels (0, 10, and \unit[20]{dB}) were adopted for the noise addition resulting in a total of 2700 tracks for each dataset. Additionally, we added a reverb with three different reverberance levels (25, 50, and \unit[100]{dB}). 

Similar to \cite{seldt}, we use an input sequence length of $512$ frames for ANSYN and MANSYN datasets while $256$ frames for REAL and MREAL datasets. To ensure a fair comparison all models were trained for $500$ epochs with the Adam \cite{adam} optimizer using the default parameters and a batch size of $16$. Early stopping is employed, where training is stopped if no improvements on validation split is observed for $50$ epochs.  

\begin{table}[t!]
\vspace{-2mm}
	\caption{\vspace{-1mm}Quantitative analysis of the complexity.\label{tab:hw}}
	\centering
	\Large
	\bgroup
	\def\arraystretch{1.1}
	\begin{adjustbox}{width=1.0\columnwidth}
		\begin{tabular}{c|cccc}
			\noalign{\smallskip} \hline \hline \noalign{\smallskip}
			Models & \# Parameters & Training time & Inference time& MACs \\
			\hline
			SELDnet& \unit[0.51]{M} & \unit[300]{s} & \unit[0.384]{s}  & \unit[1.5]{G} \\
			
			SELD-TCN& \unit[1.52]{M} & \unit[70]{s} & \unit[0.012]{s} & \unit[2]{G} \\			
			\noalign{\smallskip} \hline \noalign{\smallskip}
		\end{tabular}
	\end{adjustbox}
	\egroup
	\vspace{-8mm}
\end{table}

\section{Results and Discussion}\label{conc}
\vspace{-0.5mm}In Table \ref{tab:sampling}, we present a quantitative comparison between SELDnet and our proposed SELD-TCN across three different sampling frequencies, using both the SED evaluation metrics (i.e., F1 and ER) and the DOA estimation metrics (i.e., FR and DE). In contrast to SELDnet, our proposed approach is superior in SED performance; achieving higher F1 score and lower ER across all datasets regardless of the sampling frequency. As for the DOA estimation, SELD-TCN  achieves better FR with higher DE  ($\approx 0.5^\circ-1.7^\circ$) in the MREAL dataset. Not only does it show that our proposed method is more robust to lower sampling frequencies, but also that bidirectional RNNs can be replaced with non-causal TCNs.    

Table \ref{tab:noise} demonstrates a quantitative comparison of the SELD performance of both SELDnet and SELD-TCN under noise and reverberance. Unlike the previous experiment, both models were trained on larger number of examples due to the addition of noise and reverb using three different SNR levels. SELD-TCN outperformed the state-of-the-art SELD performance across different types of noise and reverbrance. On the other hand, SELDnet achieves slightly better DE ($\approx 0.5^\circ$) over the  ANSYN dataset. 

Table \ref{tab:hw} presents a complexity analysis in terms of the number of parameters, MAC operations, average training time per epoch and the average inference time per example. Although SELD-TCN has a higher number of parameters and MACs, it exhibits faster inference and training time due to the parallelizable nature of the convolutions. 
  
\section{Conclusion}
In this work, we have introduced a more robust and hardware-friendly SELD system. By hardware-friendly, we mean that our system can be deployed in hardware designed to execute standard CNNs. Consequently, rather than utilizing two separate hardware acceleration engines (i.e., one for CNNs and another for RNNs), the same accelerator can be used. This is favorable for embedded systems and sensor applications with limited computation resources. In order to achieve this hardware-compatibility, we have developed a temporal convolutional network based architecture, SELD-TCN, to jointly detect the start and end times of active sound events and localize each sound source. We have evaluated SELD-TCN on four different datasets featuring stationary and dynamic sound sources in anechoic and reverberant scenarios. Moreover, we have investigated the robustness to lower sampling rates, noise, and reverberance. We have shown that TCNs are able to deliver superior SELD performance and recurrent layers are not required for such a task. Furthermore, SELD-TCN was shown to be robust against low sampling rates and different types of noise and reverberance.                 


\begin{thebibliography}{10}
\providecommand{\url}[1]{#1}
\csname url@samestyle\endcsname
\providecommand{\newblock}{\relax}
\providecommand{\bibinfo}[2]{#2}
\providecommand{\BIBentrySTDinterwordspacing}{\spaceskip=0pt\relax}
\providecommand{\BIBentryALTinterwordstretchfactor}{4}
\providecommand{\BIBentryALTinterwordspacing}{\spaceskip=\fontdimen2\font plus
\BIBentryALTinterwordstretchfactor\fontdimen3\font minus
  \fontdimen4\font\relax}
\providecommand{\BIBforeignlanguage}[2]{{%
\expandafter\ifx\csname l@#1\endcsname\relax
\typeout{** WARNING: IEEEtran.bst: No hyphenation pattern has been}%
\typeout{** loaded for the language `#1'. Using the pattern for}%
\typeout{** the default language instead.}%
\else
\language=\csname l@#1\endcsname
\fi
#2}}
\providecommand{\BIBdecl}{\relax}
\BIBdecl

\bibitem{wildlife}
T.~Marques, L.~Thomas, S.~Martin, D.~Mellinger, J.~A~Ward, D.~Moretti,
  D.~Harris, and P.~Tyack, ``{Estimating animal population density using
  passive acoustics},'' \emph{Biological reviews of the Cambridge Philosophical
  Society}, vol.~88, Nov. 2012.

\bibitem{musictrans}
E.~Benetos and T.~Weyde, ``{An Efficient Temporally-Constrained Probabilistic
  Model for Multiple-Instrument Music Transcription},'' Oct. 2015.

\bibitem{musicinst}
T.~Heittola, A.~Klapuri, and T.~Virtanen, ``{Musical Instrument Recognition in
  Polyphonic Audio Using Source-Filter Model for Sound Separation.}'' Jan.
  2009, pp. 327--332.

\bibitem{autoaudioidx}
D.~Rossiter, G.~Lam, and B.~Mak, ``{Automatic Audio Indexing and Audio Playback
  Speed Control as Tools for Language Learning},'' July 2006, pp. 290--299.

\bibitem{doa1}
R.~Roden, S.~Gerlach, N.~Moritz, S.~Weinzierl, and S.~Goetze, ``{On sound
  source localization of speech signals using deep neural networks},'' March
  2015.

\bibitem{doa2}
X.~Xiao, S.~Zhao, X.~Zhong, D.~Jones, E.~Chng, and H.~Li, ``{A learning-based
  approach to direction of arrival estimation in noisy and reverberant
  environments},'' vol. 2015, pp. 2814--2818, Aug. 2015.

\bibitem{doa3}
R.~Takeda and K.~Komatani, ``{Discriminative multiple sound source localization
  based on deep neural networks using independent location model},'' Dec. 2016,
  pp. 603--609.

\bibitem{doa4}
A.~Nugraha, A.~Liutkus, and E.~Vincent, \emph{{Deep Neural Network Based
  Multichannel Audio Source Separation}}, Jan. 2017.

\bibitem{doa5}
F.~Vesperini, P.~Vecchiotti, E.~Principi, S.~Squartini, and F.~Piazza, ``{A
  neural network based algorithm for speaker localization in a multi-room
  environment},'' Sep. 2016, pp. 1--6.

\bibitem{doa6}
S.~Chakrabarty and E.~Habets, ``{Broadband DOA estimation using convolutional
  neural networks trained with noise signals},'' Oct. 2017.

\bibitem{doanet}
S.~Adavanne, A.~Politis, and T.~Virtanen, ``{Direction of Arrival Estimation
  for Multiple Sound Sources Using Convolutional Recurrent Neural Network},''
  Oct. 2017.

\bibitem{classic1}
J.~Schröder, B.~Cauchi, M.~Schädler, N.~Moritz, K.~Adiloglu, J.~Anemüller,
  S.~Doclo, B.~Kollmeier, and S.~Goetze, ``{Acoustic event detection using
  signal enhancement and spectro-temporal feature extraction},'' in \emph{Proc.
  IEEE AASP Challenge: Detection and Classification of Acoustic Scenes and
  Events}, 2013.

\bibitem{classic2}
T.~Heittola, A.~Mesaros, A.~Eronen, and T.~Virtanen, ``{Context-dependent sound
  event detection},'' \emph{EURASIP Journal on Audio, Speech, and Music
  Processing}, vol. 2013, p.~1, Jan. 2013.

\bibitem{nmf1}
T.~Heittola, A.~Mesaros, T.~Virtanen, and A.~Eronen, ``{Sound Event Detection
  in Multisource Environments Using Source Separation},'' Jan. 2011.

\bibitem{nmf2}
S.~Innami and H.~Kasai, ``{NMF-based environmental sound source separation
  using time-variant gain features},'' \emph{Computers and Mathematics with
  Applications}, vol.~64, p. 1333–1342, Sep. 2012.

\bibitem{nmf3}
A.~Dessein, A.~Cont, and G.~Lemaitre, ``{Real-Time Detection of Overlapping
  Sound Events with Non-Negative Matrix Factorization},'' \emph{Matrix
  Information Geometry}, Aug. 2013.

\bibitem{nmf4}
T.~Komatsu, T.~Toizumi, R.~Kondo, and Y.~Senda, ``{Acoustic Event Detection
  Method Using Semi-Supervised Non-Negative Matrix Factorization with Mixtures
  of Local Dictionaries},'' Sep. 2016.

\bibitem{sedcnn1}
H.~Zhang, I.~Mcloughlin, and Y.~Song, ``{Robust sound event recognition using
  convolutional neural networks},'' April 2015, pp. 559--563.

\bibitem{sedcnn2}
H.~Phan, L.~Hertel, M.~Maaß, and A.~Mertins, ``{Robust Audio Event Recognition
  with 1-Max Pooling Convolutional Neural Networks},'' Sep. 2016.

\bibitem{sedcnn3}
P.~Zinemanas, P.~Cancela, and M.~Rocamora, ``{End-to-end Convolutional Neural
  Networks for Sound Event Detection in Urban Environments},'' April 2019.

\bibitem{sedrnn1}
S.~Adavanne, G.~Parascandolo, P.~Pertilä, T.~Heittola, and T.~Virtanen,
  ``{Sound Event Detection in Multichannel Audio Using Spatial and Harmonic
  Features},'' Sep. 2016.

\bibitem{sedrnn2}
T.~Hayashi, S.~Watanabe, T.~Toda, T.~Hori, J.~Le~Roux, and K.~Takeda,
  ``{Duration-Controlled LSTM for Polyphonic Sound Event Detection},''
  \emph{IEEE/ACM Transactions on Audio Speech and Language Processing},
  vol.~25, Aug. 2017.

\bibitem{sedrnn3}
M.~Zöhrer and F.~Pernkopf, ``{Virtual Adversarial Training and Data
  Augmentation for Acoustic Event Detection with Gated Recurrent Neural
  Networks},'' Aug. 2017, pp. 493--497.

\bibitem{sedcrnn1}
E.~Cakir, G.~Parascandolo, T.~Heittola, H.~Huttunen, and T.~Virtanen,
  ``{Convolutional Recurrent Neural Networks for Polyphonic Sound Event
  Detection},'' \emph{IEEE/ACM Transactions on Audio, Speech, and Language
  Processing}, vol.~25, Feb. 2017.

\bibitem{sedcrnn2}
S.~Adavanne, P.~Pertilä, and T.~Virtanen, ``{Sound Event Detection Using
  Spatial Features and Convolutional Recurrent Neural Network},'' March 2017.

\bibitem{sedcrnn3}
S.~Adavanne, A.~Politis, and T.~Virtanen, ``{Multichannel Sound Event Detection
  Using 3D Convolutional Neural Networks for Learning Inter-channel
  Features},'' Jan. 2018.

\bibitem{sedcrnn4}
E.~Cakir and T.~Virtanen, ``{End-to-End Polyphonic Sound Event Detection Using
  Convolutional Recurrent Neural Networks with Learned Time-Frequency
  Representation Input},'' July 2018, pp. 1--7.

\bibitem{sedcrnn5}
H.~Phan, O.~Chén, P.~Koch, L.~Pham, I.~Mcloughlin, A.~Mertins, and M.~de~Vos,
  ``{Unifying Isolated and Overlapping Audio Event Detection with Multi-Label
  Multi-Task Convolutional Recurrent Neural Networks},'' May 2019.

\bibitem{sedcrnn6}
S.~Jung, J.~Park, and S.~Lee, ``{Polyphonic Sound Event Detection Using
  Convolutional Bidirectional LSTM and Synthetic Data-based Transfer
  Learning},'' May 2019, pp. 885--889.

\bibitem{rnnhw1}
A.~Chang, B.~Martini, and E.~Culurciello, ``{Recurrent Neural Networks Hardware
  Implementation on FPGA},'' Nov. 2016, p.~9.

\bibitem{rnnhw2}
A.~Chang and E.~Culurciello, ``{Hardware accelerators for recurrent neural
  networks on FPGA},'' May 2017, pp. 1--4.

\bibitem{cnns}
J.~Gu, Z.~Wang, J.~Kuen, L.~Ma, A.~Shahroudy, B.~Shuai, T.~Liu, X.~Wang, and
  G.~Wang, ``{Recent Advances in Convolutional Neural Networks},''
  \emph{Pattern Recognition}, Dec. 2015.

\bibitem{wavetcn}
\BIBentryALTinterwordspacing
A.~van~den Oord, S.~Dieleman, H.~Zen, K.~Simonyan, O.~Vinyals, A.~Graves,
  N.~Kalchbrenner, A.~Senior, and K.~Kavukcuoglu, ``{WaveNet A Generative Model
  for Raw Audio},'' 2016. [Online]. Available:
  \url{https://arxiv.org/abs/1609.03499}
\BIBentrySTDinterwordspacing

\bibitem{tcn}
S.~Bai, J.~Kolter, and V.~Koltun, ``{An Empirical Evaluation of Generic
  Convolutional and Recurrent Networks for Sequence Modeling},'' March 2018.

\bibitem{seldt}
S.~Adavanne, A.~Politis, and T.~Virtanen, ``{Localization, Detection and
  Tracking of Multiple Moving Sound Sources with a Convolutional Recurrent
  Neural Network},'' April 2019.

\bibitem{batchnorm}
S.~Ioffe and C.~Szegedy, ``{Batch Normalization: Accelerating Deep Network
  Training by Reducing Internal Covariate Shift},'' Feb. 2015.

\bibitem{wavenetdenoise}
D.~Rethage, J.~Pons, and X.~Serra, ``{A Wavenet for Speech Denoising},'' June
  2017.

\bibitem{hwdnnsurvey}
V.~Sze, Y.-H. Chen, T.-J. Yang, and J.~Emer, ``{Efficient Processing of Deep
  Neural Networks: A Tutorial and Survey},'' \emph{Proceedings of the IEEE},
  vol. 105, March 2017.

\bibitem{dataset}
\BIBentryALTinterwordspacing
S.~Adavanne, J.~Nikunen, A.~Politis, and T.~Virtanen, ``{TUT Sound Events 2018
  - Ambisonic, Reverberant and Real-life Impulse Response Dataset},'' April
  2018. [Online]. Available: \url{https://doi.org/10.5281/zenodo.1237793}
\BIBentrySTDinterwordspacing

\bibitem{seldnet}
S.~Adavanne, A.~Politis, and T.~Virtanen, ``{Direction of Arrival Estimation
  for Multiple Sound Sources Using Convolutional Recurrent Neural Network},''
  Oct. 2017.

\bibitem{demand}
J.~Thiemann, N.~Ito, and E.~Vincent, ``{The Diverse Environments Multi-Channel
  Acoustic Noise Database (DEMAND): A database of multichannel environmental
  noise recordings},'' \emph{The Journal of the Acoustical Society of America},
  vol. 133, p. 3591, May 2013.

\bibitem{adam}
D.~Kingma and J.~Ba, ``{Adam: A Method for Stochastic Optimization},''
  \emph{International Conference on Learning Representations}, Dec. 2014.

\end{thebibliography}
\begin{spacing}{0.9185}
	\bibliographystyle{IEEEtran}

\end{spacing}

\end{document}